\documentclass[pre%
,twocolumn%
,superscriptaddress%
,floats
,aps%
,amssymb%
]{revtex4}
\usepackage{amsmath}%

\begin{document}
\title{Comparing extremal and thermal Explorations of Energy Landscapes} 
\date{\today}
\author{Stefan Boettcher}  
\email{www.physics.emory.edu/faculty/boettcher}
\affiliation{Physics Department, Emory University, Atlanta, Georgia
30322, USA} 
\author{Paolo Sibani}  
\email{Permanent Address: Fysisk Institut, SDU, Odense, DK}
\affiliation{Theoretical Physics, Oxford University, 1 Keble Rd, Oxford OX1 3NP, UK}   

\begin{abstract} 
Using a non-thermal local search, called Extremal Optimization (EO),
in conjunction with a recently developed scheme for classifying the
valley structure of complex systems, we analyze a short-range spin
glass.  In comparison with earlier studies using a thermal algorithm
with detailed balance, we determine which features of the landscape
are algorithm dependent and which are inherently geometrical.
Apparently a characteristic for any local search in complex energy
landscapes, the time series of successive energy records found by EO
also is characterized approximately by a log-Poisson statistics.
Differences in the results provide additional insights into the
performance of EO.  In contrast with a thermal search, the extremal
search visits dramatically higher energies while returning to more
widely separated low-energy configurations. Two important properties
of the energy landscape are independent of either algorithm: first, to
find lower energy records, progressively higher energy barriers need
to be overcome. Second, the Hamming distance between two consecutive
low-energy records is linearly related to the height of the
intervening barrier.
\hfil\break  PACS number(s): 
05.40.-a
, 75.10.Nr
, 02.60.Pn
.
\end{abstract} 
\maketitle

\section{Introduction}
\label{intro}
The exploration of complex energy landscapes poses a series of
problems of wide interest. Their multi-modal geometry is on one side
challenging for optimization algorithms attempting to find the global
minimum~\cite{Hartmann04}, while on the other side it provides a
framework to model slow relaxation dynamics in
nature~\cite{Frauenfelder96}. Typically, in a physical (thermal)
exploration of such relaxation phenomena, the static (geometric) and
dynamic (algorithmic) aspects are intertwined. To disentangle the inherently
geometric features, we apply a decidedly non-thermal optimization
algorithm, called Extremal Optimization
(EO)~\cite{Boettcher00,Boettcher01a,Hartmann04}, to explore the energy
landscape of a spin glass whose structure has been studied recently
with a thermal algorithm~\cite{Dall03}.  {}Furthermore, the comparison
with the thermal algorithm highlights distinct performance features of
the EO algorithm.

{}Focusing on the temporal succession of energy values of record
magnitude, we present a set of measures which characterizes the
difficulty of local searches and the complexity of the
landscape. There are at least two geometrical features of the spin
glass landscape which are robust. For one, progressively higher energy
states have to be surmounted in order to reach ever lower energy
records. Second, a linear relation emerges between the Hamming
distance of consecutive low-energy record configurations and the
height of the highest intervening energy state. Such a relation has
previously been found by other authors using different local search
methods, as well as for other models~\cite{Nemoto88,Vertechi89,Billoire01}.
Yet, in stark contrast to the thermal method, EO reaches a rapidly
(exponentially) growing succession of high-energy records, projecting
the search through configuration space by an exponentially growing
Hamming distance between consecutive low-energy records.

In the following section, we describe how the landscape features
explored by the dynamics can be assessed on the basis of a time series
of suitably defined `valleys'. In Sec.~\ref{eo}, we review the
Extremal Optimization heuristic used here to produce such a time
series for a non-thermal relaxation process. The main results of our
numerical studies are discussed in Sec.~\ref{results}, and in
Sec.~\ref{conclusions} we present our conclusions.

\section{Energy Valleys in Complex Landscapes}
\label{energy_valleys} 
The idea that progressively deeper, i.e.\ thermally more stable
valleys are explored by the thermal dynamics of complex system is
well-established and accounts for many important features of aging
dynamics~\cite{Sibani89,Lederman91,Sibani91,Bouchaud95,Joh96,Hoffmann97,Sibani97a,Joh99,Crisanti00,Buisson03}.
Qualitatively speaking, a valley would be a set of configurations
sufficiently close to a local energy minimum. These configurations are
repeatedly visited by the dynamics before the valley is ``exited'' and
altogether different regions are explored. Ideally, in thermal systems
the states within a valley would be visited with frequencies given by
the respective Boltzmann weights, i.e.\ a local thermal equilibrium
state is established before the valley is left.

\begin{figure}
\vskip 2.5in \includegraphics{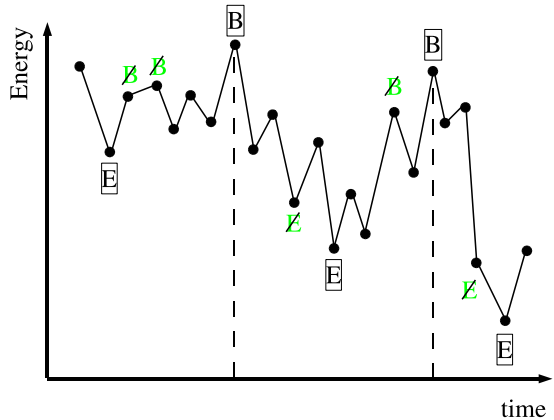} 
\caption{The definition of a valley is illustrated with a fictitious
series of energy values. A search produces a time sequence of energy
(E) and barrier (B) records, where each ``E'' labels the lowest energy
seen so far, and ``B'' refers to the highest barrier ({\it relative to
the most recent ``E''}) reached up to that time. As explained in the
text, in the end, only the highest barriers and the lowest energy
records in each subsequence of ``E''s and ``B''s is kept, and the
intermediate values are stricken from the record to give a strictly
alternating sequence ``EBEBE...''. In particular, any two subsequent
``B''s demarcate (entrance to and exit from) valleys, here separated
by vertical dashed lines.  }
\label{land_sketch}
\end{figure} 

\begin{figure}
\vskip 4.85in \includegraphics{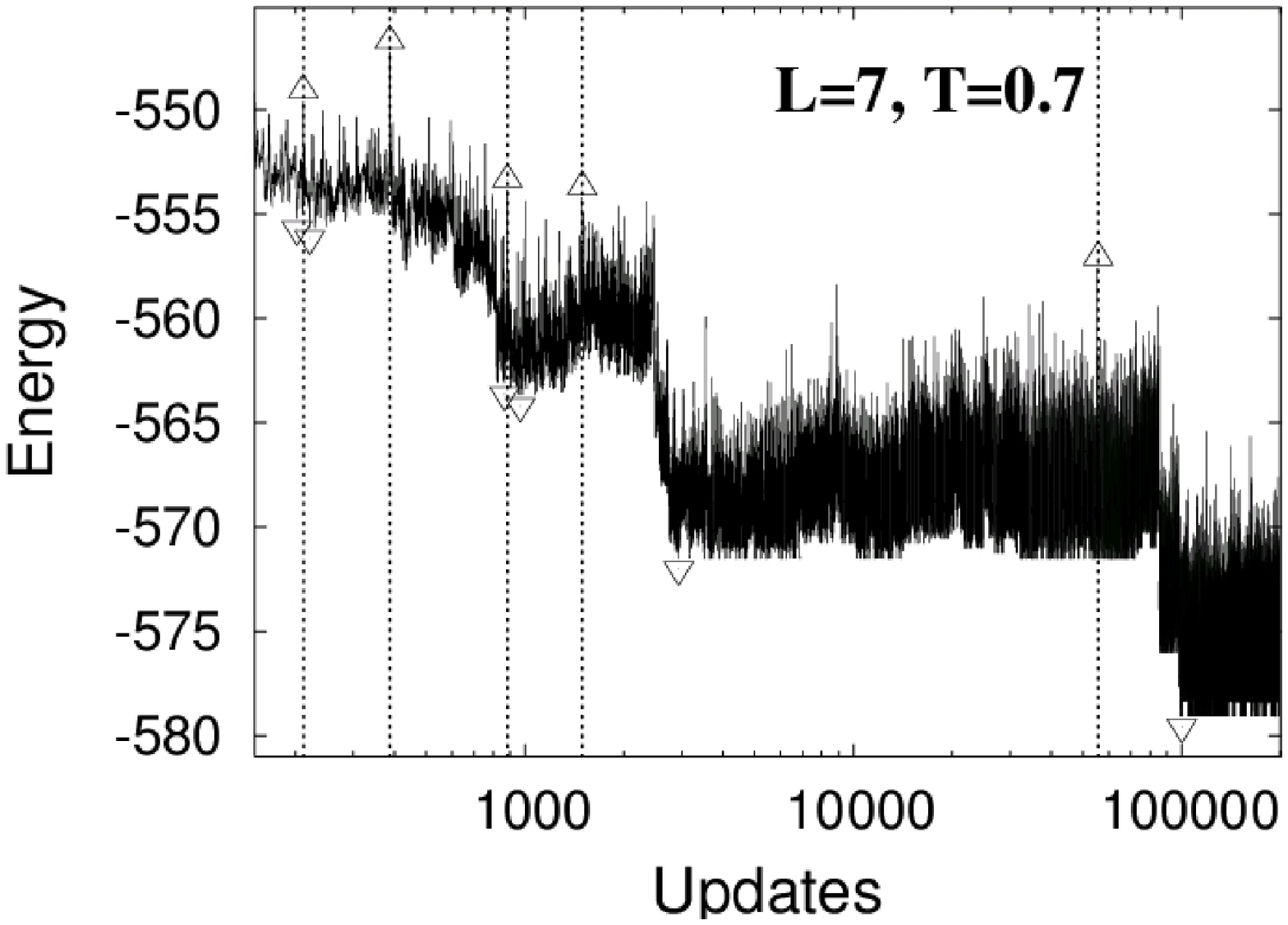} 
\includegraphics{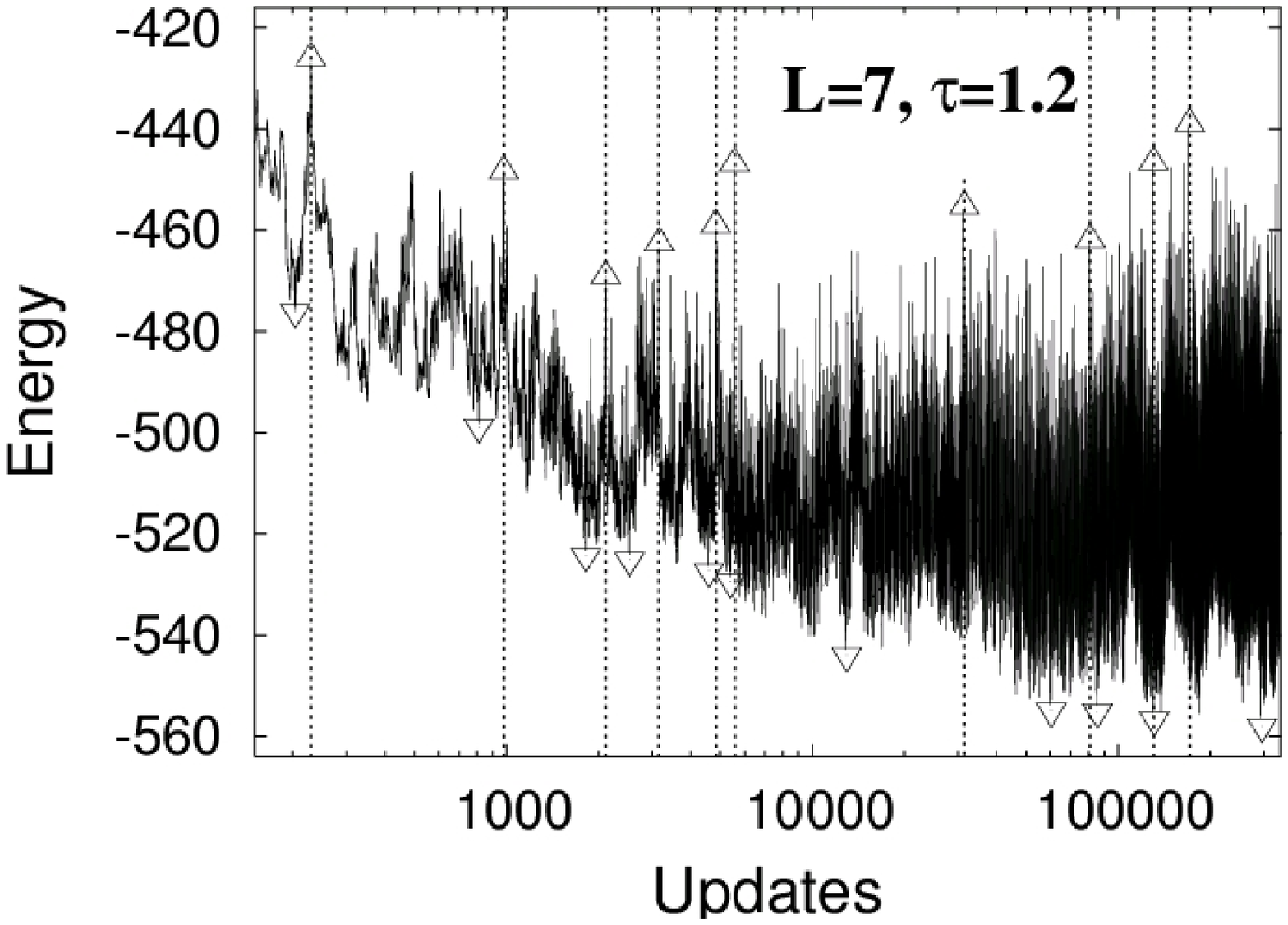}
\caption{Plot of a typical run with a thermal search (above) and an
extremal search (below) for a $d=3$ Gaussian spin glass of size
$L=7$. The fluctuating line marks the sequence of energies visited by
the search. In terms of the definition in the text and
Fig.~\protect\ref{land_sketch}, energy records (E) are marked by
down-triangles, barrier records (B) by up-triangles.  The barrier
records also demarcate the beginning and the end of a valley, so each
time interval between two consecutive vertical lines constitutes a
valley. Counting valleys starts (with $n_V=0$) for updates $>N$ (where
$N=7^3=343$ here) to avoid early transient behavior. While the
absolute energy scale between both searches is not significant here
(two distinct bond realizations were used), the difference in range
and shape of the fluctuations is remarkable.}
\label{valleyplot}
\end{figure}

The standard way of giving these concepts an operational meaning is to
subject the dynamics to repeated thermal quenches, each leading to a
local energy minimum or Intrinsic State (IS). The set of IS thus
obtained partitions configuration space into basins of attraction with
respect to thermal quenches, and each basin can be considered as a
valley.  A different approach~\cite{Dall03}, which we presently
generalize to non-thermal algorithms such as EO, uses a time series of
unperturbed energy data.  It defines valleys on the basis of the
lowest energy visited {\it so far} and is motivated by the fact that,
for thermal dynamics, the dynamics is recurrent and equilibrium-like
as long as this lowest energy state does not change.

As the details can be found in Ref.~\cite{Dall03}, only a brief
account will be given for completeness (see Fig.~\ref{land_sketch}):
We keep track of the current lowest energy value $E$ encountered up to
time $t$ and always measure the energy of the current state as the
difference from this particular value. We furthermore keep track of
the highest energy barrier $B$ visited, of the times at which both low
and high records occur, and of the corresponding configurations.  A
trajectory is thus mapped into a symbolic sequence such as $\ldots
EEEBBBBBEEEEEEBBEEE\ldots$, where the number of symbols in each
subsequence of contiguous $E$ or $B$ values is larger or equal to one.

Consider first any subsequence of $E$'s: since a trajectory `sliding'
downhill will produce such a sequence, all but the last $E$ correspond
to transient states. By contrast, the last $E$ record may stand for a
long time, i.e.\ at least until a record high energy value, the first
subsequent $B$, is encountered. This makes it a good proxy for the
lowest energy value in the `current' valley.  Similarly, while a
trajectory explores high energy states it will likely visit several
closely spaced energy maxima, producing a subsequence of $B$
values. The last $B$ value remains a record for long time, and is
chosen to mark the passage from one valley to the next.  In summary,
all finite subsequences are pruned to their last element, producing a
sequence $\ldots BEBEBE \dots$ where each triplet $BEB$ marks a
valley, see Fig.~\ref{land_sketch}. The $B$s mark the enter and exit
events and the the $E$ marks the lowest energy of the valley. The last
sequence of $E$'s or $B$'s encountered in the simulation is not
terminated and is discarded.

We note that this scheme produces trivial results (very few or no
valleys, only infinitesimally increasing barriers, nearly same
configurations for different $E$ 's) in cases where there is only one
global minimum, or when the ground state can be chosen as starting
point.  Consider e.g. the case of a Metropolis random walk in a
discrete set of energy value with a single minimum at energy zero. The
downward drift implies that this minimum would most likely be reached
without intervening barrier maxima, and the scheme will produce no
output. In cases where a few `false positives' are produced,
i.e. valleys with no physical counterparts, these will closely
resemble each other in terms of their configurations, energies
etc. and will be immediately identified as such in the subsequent
analysis.  In any case, the time needed to reach the global minimum
depends linearly, or in the lack of a bias, quadratically on the
initial energy.  The situation is completely different in complex
energy landscapes, where, for thermal dynamics~\cite{Dall03}, and as
shown below, for EO as well, the scheme gives a succinct but highly
informative description of the dynamics: new valleys are accessed on a
logarithmic time scale, and there is a systematic variation of their
properties with the 'valley index'.  We finally note that subsequent
valley are defined on coarsening energy and time scales. Hence, each
valley can be expected to contain many valleys of the previous kinds
as sub-features.  An illustration of the valley structure in a thermal
and an extremal search is given in Fig.~\ref{valleyplot}.

{}From an optimization point of view, the $E$ values in the series
represent the best the algorithm can do on a given time scale.  The
$B$ values act as energy barriers for a thermal type of algorithm, but
not for an algorithm of the EO type, where energy differences have
less dynamical significance. Nevertheless, if, as we expect, a
geometric relationship links the Hamming distance between two
``sufficiently low'' minima and the height of the intervening barrier,
the link should appear irrespective of the algorithm chosen.  As shown
below, the expectation is borne out by our simulations.

\section{Extremal Optimization}
\label{eo}
An energy record statistic as described in the previous section can be
generated by a variety of dynamical rules. Physically most relevant
are those which evolve according to a thermal process that preserves
detailed balance. The record statistics of a thermal process has been
extensively studied previously for a number of different
systems~\cite{Dall03}, including the Edwards-Anderson model with
Gaussian couplings on a cubic lattice. A priori, it is not obvious
which of the properties of this process can be attributed to the
dynamic update rule, and which are inherently properties of the
system.

A significant alteration of the update dynamics may in turn elucidate
the origin of certain properties. To this end, we consider the
Extremal Optimization (EO) heuristic~\cite{Boettcher00,Boettcher01a}
as a decidedly different, athermal, update rule to explore the
system. EO, like Simulated Annealing
(SA)~\cite{Kirkpatrick83,Salamon02}, attempts to advance toward lower
energy values via a local search of the landscape. Unlike SA, EO is
modeled after driven dissipative processes, intentionally pushing the
dynamics away from local equilibrium and detailed balance.
 
While a distinction between static landscape and dynamic properties is
desirable in its own right, the similarity of these physical systems
to many practical combinatorial optimization problems in computer
science provides additional incentive for a broad-based investigation
of local search methods and their ability to exploit the landscape
geometry. EO in particular has proved to be a competitive heuristic to
determine low-energy configuration for some of the hardest
combinatorial optimization problems known, including graph
bipartitioning~\cite{Boettcher99},
coloring~\cite{Boettcher01a,Boettcher04}, and also spin glass
problems~\cite{Boettcher03}. We can hope that a more detailed view at
the interplay of heuristic search and landscape geometry will lead to
improvements in the quality of the results found as well as in the
speed of convergence.

The extremal optimization algorithm, $\tau$-EO, which we employ in
this paper, has been discussed previously in \cite{Boettcher01a}, and
in \cite{Boettcher02,Boettcher01b} with regard to the setting of its
sole free parameter, $\tau$. At each instant during the search of a
particular instance, $\tau$-EO assigns to each spin $x_i$ in the
configuration its contribution to the total energy as ``fitness,''
\begin{eqnarray}
\lambda_i=\frac{1}{2}x_i\sum_{<,j>}J_{i,j}\,
x_j-\frac{1}{2}\sum_{<,j>}\left|J_{i,j}\right|,
\label{lambdaeq}
\end{eqnarray}
where the summation extends over all neighboring spins $x_j$ of
$x_i$. Note that the second term on the right corresponds to the
(absolute) weight attributable to that spin; it ensures that for each
variable its optimal fitness is zero, irrespective of its overall
weight. Accordingly,
\begin{eqnarray}
H=-\sum_i\lambda_i-{\sum\sum}_{<i,j>}\left|J_{i,j}\right|,
\label{Heq}
\end{eqnarray}
i.~e. the sum of all fitnesses tallies the total energy, aside from a
trivial offset.

During a search with $\tau$-EO, we rank all $x_i$ according to fitness
$\lambda_i$, {\em i.e.\/}, we find a permutation $\Pi$ of the
variable labels $i$ with
\begin{eqnarray}
\lambda_{\Pi(1)}\leq\lambda_{\Pi(2)}\leq\ldots\leq\lambda_{\Pi(n)}.
\label{rankeq}
\end{eqnarray}
The variable $x_j$ with the worst burden on the total energy is of
rank 1, $j=\Pi(1)$, and the best variable is of rank $n$. Consider a
scale-free probability distribution over the {\em ranks\/} $k$,
\begin{eqnarray}
P_k\propto k^{-\tau},\qquad 1\leq k\leq n,
\label{taueq}
\end{eqnarray}
for a fixed value of $\tau$. At each update, select a rank $k$
according to $P_k$.  The spin $x_j$ with $j=\Pi(k)$ is forced {\it
unconditionally} to change state.  For $\tau>0$ this selection process
ensures a certain preference in fixing the state of spin variables
which put a higher burden on the total energy. In particular, it has
been found that intermediate choices for the value of $\tau$ in
Eq.~(\ref{taueq}), with $\tau-1\sim 1/\ln(n)$~\cite{Boettcher02}, often
lead to the best results for a given runtime of the algorithm. Values
of $\tau$ much larger or smaller than these quickly produce  too
confined or too random searches, as we will see. This issue has been
explored also in Ref.~\cite{Middleton04}.

The definition of fitness, which generally permits variations that can
significantly impact performance~\cite{Boettcher00}, as given in
Eq.~(\ref{lambdaeq}) is purely a measure of ``badness'' in each
variable. The constant offset in each $\lambda_i$ in Eq.~(\ref{lambdaeq}),
consisting of the absolute weight of all attached bonds, ensures that
perfectly well adapted spins, {\it i.~e.} those bordering on bonds that are
all satisfied, have zero fitness, irrespective of their overall
weight. Spins bordering on unsatisfied bonds get penalized according
to their burden on the total energy. Considering that the absolute sum
of all weights associated with spins is distributed unevenly,
heavy-weighted spins (and bonds) are satisfied with a higher
priority. At later stages of the search, the overwhelming number of
weaker bonds are attended to.

\section{Numerical Results}
\label{results}
In this section, we present the results of extensive numerical
investigations of the $\tau$-EO search for the $d=3$ Edwards-Anderson
spin glass. The procedure, as outlined in Sec.~\ref{energy_valleys},
follows closely that of Ref.~\cite{Dall03}. There, a rejection free
implementation of the Metropolis algorithm, the Waiting Time Method
(WTM)~\cite{Dall01}, was used to study the thermal dynamics of this
spin glass at low temperatures starting from a hard quench. In the
process, many salient features of the record statistics have been
measured, which we will focus on also in this study.

We run the $\tau$-EO search for a large number of instances of varying
lattice size to explore the finite-size scaling properties of the
observables with sufficient accuracy. In particular, we have used from
100,000 bond realizations for $L=8$ to 1,000 such instances for $L=20$
and performed 3 runs each.  As in the thermal process, each run starts
from random initial conditions in the spin configuration. A form of
local ``equilibration'' is reached within a few EO-sweeps of the
system, during which each spin gets to arrange itself with the local
field imposed by its neighbors. After one sweep, we start sampling
low-energy and barrier records through a sequence of valleys. In
particular, we run the EO algorithm on each instance for $O(N^2)$
update steps, where $N=L^3$, or about $10N$ sweeps of the
system. Here, a ``sweep'' refers to $N$ EO-update steps, which are
stochastic and do not imply that each variable is updated exactly
once.  Note that the valley index is gauged to be $n_V=0$ after the
first ($t=1$) sweep of the system.

\subsection{Varying $\tau$}
\label{tauvary}
First, we have studied the dependence of EO's performance on the
parameter $\tau$. To this end, we have conducted about 10,000 runs of
EO on random instances at fixed system size $L=16$ for $\tau=0.2$,
0.7, 1.2, 1.7, and 2.2. The behavior of EO has previously been show to
be very sensitive to this parameter. Overall, we notice that, indeed,
the results of the record statistics are as well strongly
$\tau$-dependent. This dependence is often not monotone in
$\tau$. 

\begin{figure}
\vskip 2.3in \includegraphics{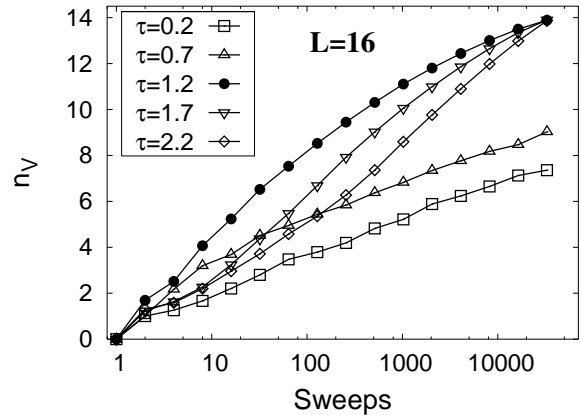}
\caption{Plot of the valleys found by EO on a logarithmic timescale
for a system of $L=16$ but with values of $\tau=0.2$, 0.7, 1.2, 1.7,
and 2.2. The exploration of new valleys occurs at a much faster rate
for intermediate values of $\tau$, until at later times saturation due
to finite size effects permits runs with higher $\tau$ to catch up.}
\label{tau_nV}
\end{figure}

Most significantly, as Fig.~\ref{tau_nV} shows, the most
valleys, and the best energy records, are obtained at
intermediate values of $\tau$, i.~e. those valleys are found on much
shorter timescales than for $\tau$ values that are too large or too
small. Since there is a rapid gain in new valleys for $\tau=1.2$,
saturation effects due to finite system size set  in faster, a trend
to which higher values of $\tau$ eventually catch up. Overall, there
is a more pronounced variation with the parameter $\tau$ than in the
corresponding data for a thermal search (see Fig.~1 in
Ref.~\cite{Dall03}). From an optimization standpoint, $\tau$-EO
progresses through these valleys with about ten times fewer sweeps
than a fixed-temperature search.

\begin{figure}
\vskip 2.3in \includegraphics{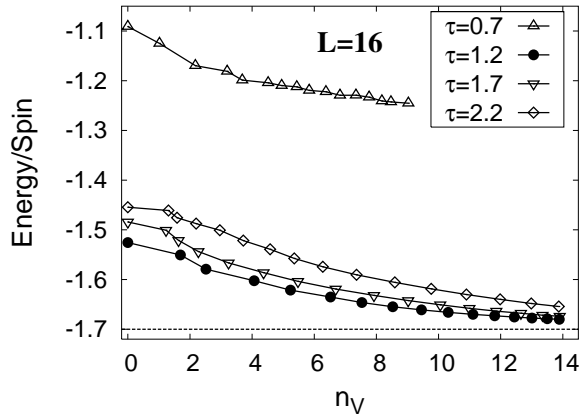}
\caption{Plot of the energy records found by EO within a given valley
as a function of the valley index $n_V$. Not only do we find new
valleys faster at intermediate $\tau\approx1.2$ (as
Fig.~\protect\ref{tau_nV} shows), but also are lower energy records
found within, even at equalized valley index $n_V$. (For $n\to\infty$,
average ground state energies are $\approx-1.70$~\protect\cite{Pal96}.)
Note the dramatic shift between $\tau=0.7$ and $\tau=1.2$. (Data for
$\tau=0.2$ is even further above that for $\tau=0.7$ and has been
dropped here.)}
\label{tau_energy}
\end{figure}

The trend that an intermediate value of $\tau$ provides the favorable
search results is also reflected in the energy records found in
those valleys, as Fig.~\ref{tau_energy} shows. Thus, for an
``optimal'' value of $\tau$ near unity, EO finds new
valleys faster {\it and} the energy states accessed within those
valleys are lower. In fact, the exploration for larger values of
$\tau$ proves qualitatively similar, just on a slower timescale. The
behavior for $\tau<1$ is quite distinct, more akin to a
high-temperature thermal diffusion. These observations are consistent
with the phase transition at $\tau=1$ in the search dynamics of
$\tau$-EO found for a model problem in Ref.~\cite{Boettcher02}.

\begin{figure}
\vskip 4.5in \includegraphics{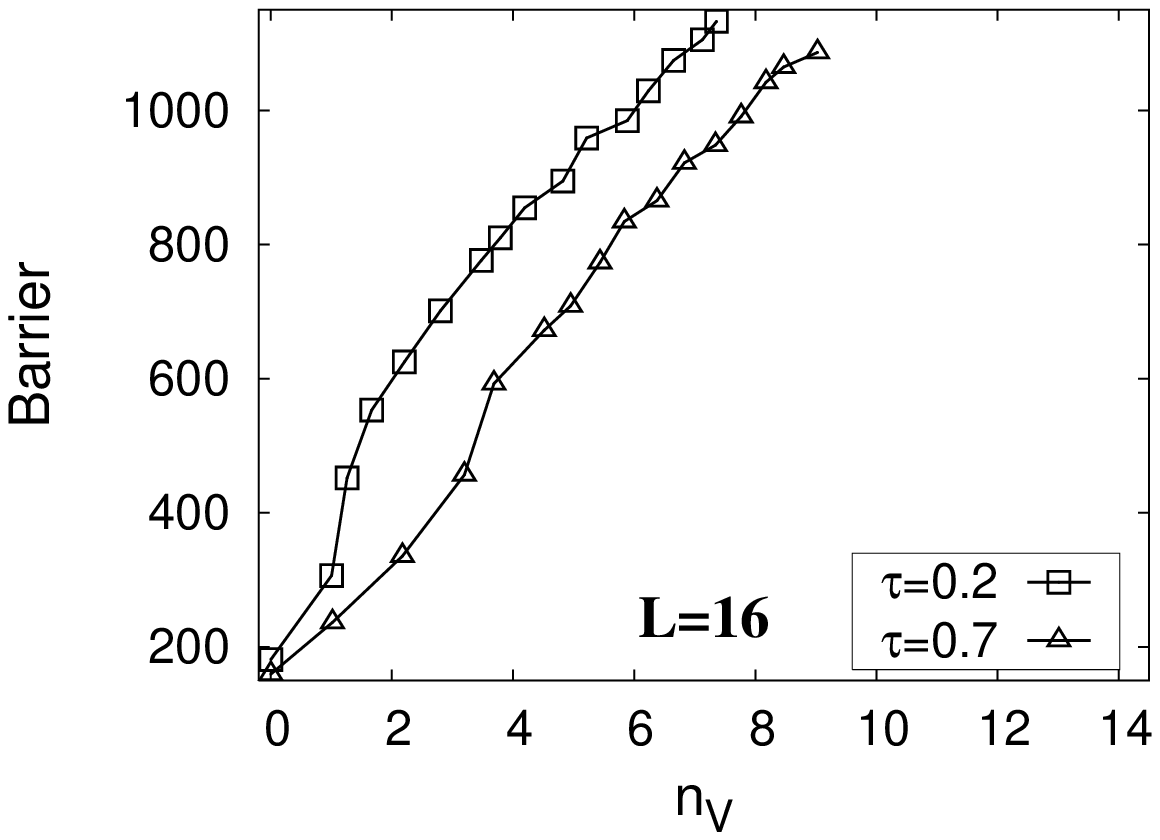}
\includegraphics{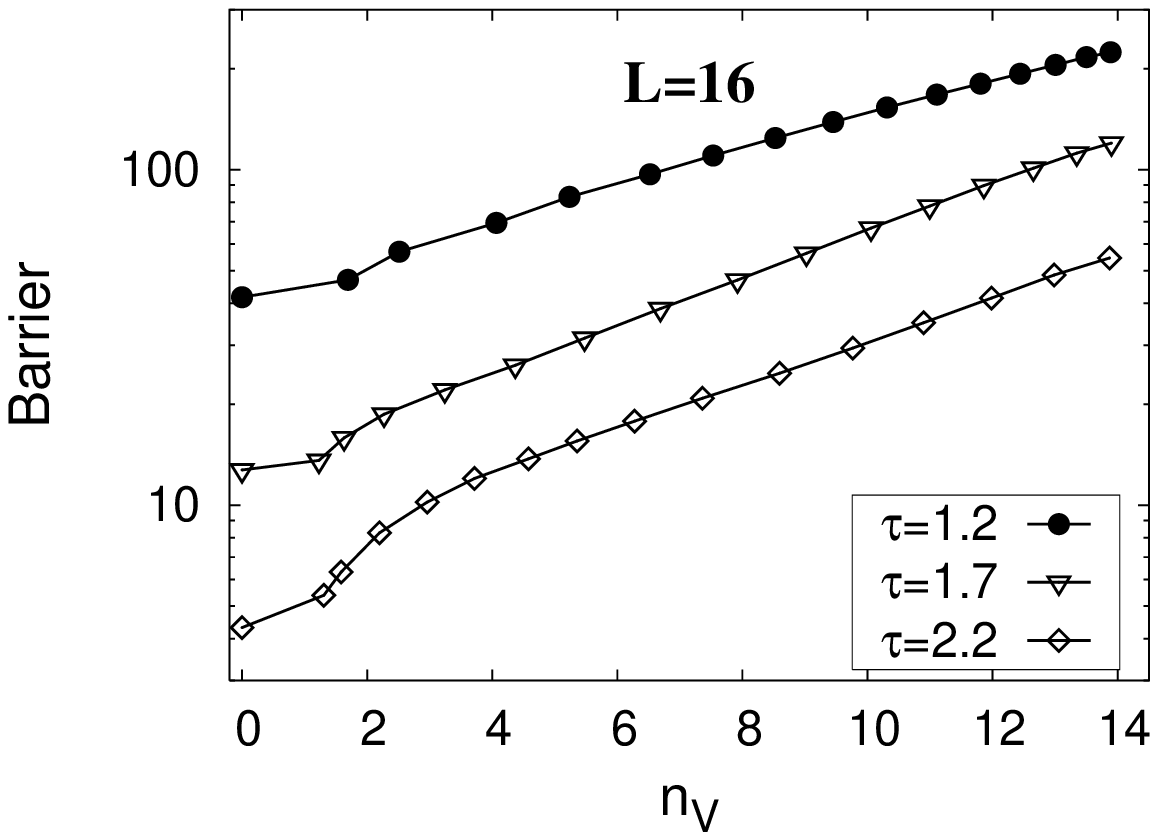}
\caption{Plot of the highest energy level (or barrier) accessed
between two successive low-energy records as a function of valley
index $n_V$. (Note that the height is measured relative to the most
recent low-energy record.) Unlike in Figs.~\protect\ref{tau_nV}
and~\protect\ref{tau_energy}, the $\tau$-dependence here is
monotone. Yet, for $\tau<1$ (upper Figure) the barrier height varies
only linearly with the valley index, similar to a thermal search,
while for $\tau>1$ (lower Figure) barrier heights rise
exponentially. }
\label{tau_barrier}
\end{figure}

\begin{figure}
\vskip 2.3in \includegraphics{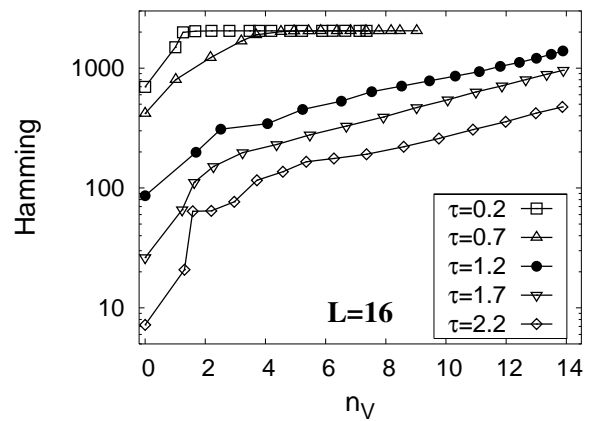}
\caption{Plot of the Hamming distance between successive low-energy
records as a function of valley index $n_V$. As in
Fig.~\protect\ref{tau_barrier}, the $\tau$-dependence here is
monotone, showing that the search gets increasingly narrow for
increasing $\tau$. For small $\tau$, the Hamming distance quickly
saturates at the system size ($n/2$).}
\label{tau_hamming}
\end{figure}

At low values of $\tau$ the search is too random, and whatever valleys
are found are not explored with sufficient ``greed.'' This randomness
expresses itself also in very high energy states accessed in between
valleys, and the total decorrelation between successive valleys as
illustrated by the large Hamming distances between the consecutive
records in energy.  (Hamming distance refers to the number of spin
flips that separates any two configurations.)  These features are
displayed in Figs.~\ref{tau_barrier} and~\ref{tau_hamming}, in which
the barrier records and Hamming distances of consecutive energy
records are plotted for each valley index. These properties are quite
monotone in $\tau$ as can be expected from the nature of the optimal
EO search as a compromise between too random and too greedy
behavior~\cite{Boettcher02,Boettcher01b}. Hence, on the greedy side,
for larger $\tau$, EO spends a long time in each valley, while
reaching down to very low-energy states within each valley, before
escaping through an equally low barrier, which only provides access to
a new valley with states highly correlated (small Hamming distance)
with those in the previous one. Conversely, for smaller $\tau$, the
search approaches a random walk through the configuration space that
is unlikely to reach down to very low-energy states.  Any memory of
fit variables in the list in Eq.~(\ref{rankeq}) is short-lived, and
the search trajectory quickly decorrelates such that the Hamming
distances between consecutive barrier records soon saturate at the
system size ($n/2$), see Fig.~\ref{tau_hamming}.

Most surprising is the rapid increase of both, barrier heights and
Hamming distances, which scale {\it exponentially} with the valley
index for sufficiently large values of $\tau$. This is in marked
difference with the behavior of the thermal relaxation, where both
grow about linearly with the index (see Figs.~2 and~3 in
Ref.~\cite{Dall03}). In contrast, near $\tau=0$ (which rigorously
corresponds to $T=\infty$) we recover the linear scaling of the
barrier heights with valley index observed for a thermal search, as
the upper plot in Fig.~\ref{tau_barrier} demonstrates, again a hint of
the phase transition at $\tau=1$. In general, barrier heights and
Hamming distances vary significantly for both, extremal and thermal
exploration, with their respective parameters.

\begin{figure}
\vskip 2.3in \includegraphics{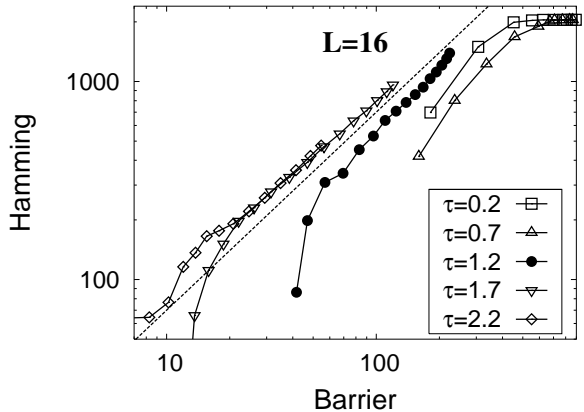}
\caption{Plot of the Hamming distance between successive low-energy
records as a function of the intervening barrier height, obtained from
Figs.~\protect\ref{tau_barrier} and~\protect\ref{tau_hamming} by
eliminating the valley index dependence between them. The relation for
each value of $\tau$ appears to be in fact linear, even for $\tau<1$
before the Hamming distances saturate. Linearity is exemplified by
the dashed line of slope 1; the log-log scale was merely chosen for
better visibility. }
\label{tau_HamBar}
\end{figure}

Despite this difference between extremal and thermal exploration, in
both cases the scaling of the Hamming distance itself with the
barrier, as shown in Fig.~\ref{tau_HamBar}, is quite linear aside from
finite-size effects. Hence, this is all but the first indication of a
quantity attributable the landscape geometry itself. The purely
geometric origin of this feature is further emphasized by the fact
that the data for all $\tau$ very nearly collapses onto a single line,
showing only a weak $\tau$-dependence in the slope. While it is not
too surprising to obtain such a linear relation from the ratio of two
linear relations for the thermal search in Ref.~\cite{Dall03}, in
turn, it is outright amazing to extract a simple linear relation from
the ratio of two exponentials in the case of EO here. A linear
relationship between Hamming distance and barrier records was obtained
long ago for the SK-model~\cite{Nemoto88,Vertechi89}, and was inferred
experimentally from thermo-remanent magnetization
data~\cite{Lederman91}.  We also note that the \emph{largest} Hamming
distance achievable for a given barrier.  as opposed to the typical
one, grows exponentially with the barrier in both 2d and 3d
spin-glasses~\cite{Sibani98}.

\begin{figure}
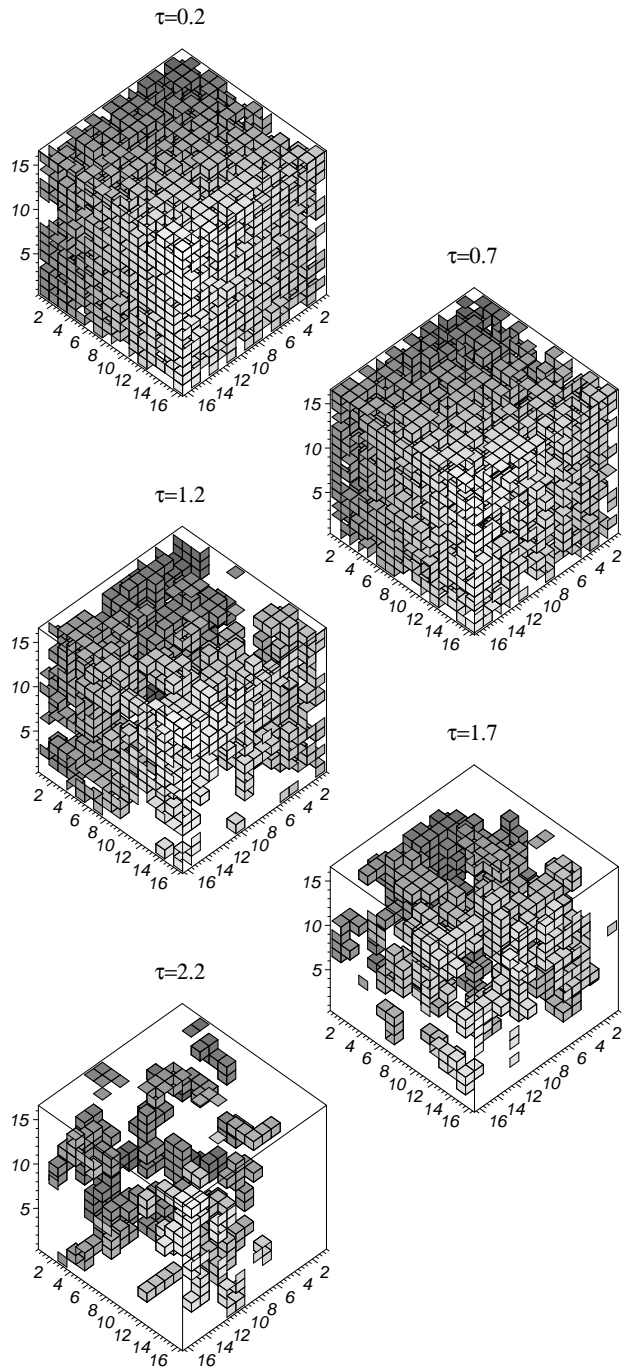

\vskip 7.5in 
\includegraphics{cluster016_020_BW.ps}
\includegraphics{cluster016_070_BW.ps}
\includegraphics{cluster016_120_BW.ps}
\includegraphics{cluster016_170_BW.ps}
\includegraphics{cluster016_220_BW.ps}
\caption{Snapshot of the clusters of spins which have changed between
  the configurations of two consecutive energy records at $n_V=9$ and
  10 found by $\tau$-EO with $\tau=0.2,\ldots,2.2$, for some instance
  of $L=16$. Patches cut across bonds along which the relative state of
  adjacent spins has changed. All surfaces are actually closed and
  only appear to be open due to the periodic boundary conditions. For
  small $\tau$, all spins appear to have flipped many times, and it is
  impossible to tell ``inside'' from ``outside,'' while for larger
  $\tau$ a large number of spins clearly  have remained ``frozen''. It
  appears that for the optimal $\tau=1.2$ flipped spins just about percolate.}
\label{tau_cluster}
\end{figure}

A qualitative snapshot of the difference between consecutive record
configurations is provided in Fig~\ref{tau_cluster}. There we plot
the clusters of overturned spins between two  energy records at
$n_V=9$ and $n_V=10$ for some random instance of $L=16$ and various
$\tau$. As can be expected from  Fig.~\ref{tau_hamming}, these records
differ by fewer spins for increasing $\tau$. Hence, for low $\tau$,
the interfaces between flipped and unflipped spins percolates and is
rather indistinct, for $\tau>1$ isolated individual clusters become
discernible.

\begin{figure}
\vskip 2.3in \includegraphics{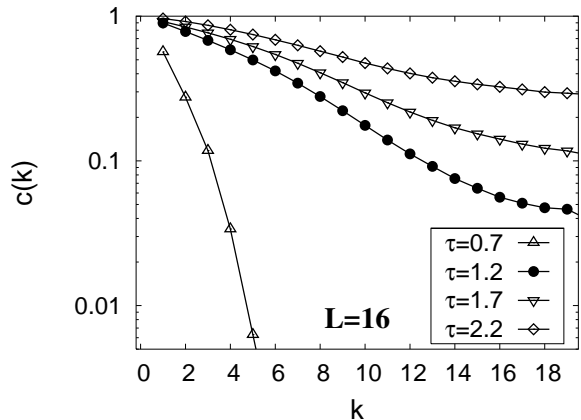}
\caption{Plot of the correlations $c(k)$ defined in
  Eq.~\protect\ref{correq} between configurations of record energy in
  valley $n_V=k$ and $n_V=0$. As in Fig.~\protect\ref{tau_hamming},
  the $\tau$-dependence here is monotone. The search for small $\tau$
  quickly decorrelates, but reaches a plateau value for $\tau>1$.}
\label{tau_corr}
\end{figure}

Further quantitative insight into the scope of the search can be
obtained by looking at the overlap between earlier-found
configurations and those arrived at for later times of the search. To
this end, we store the optimal configuration found after one sweep in
valley $n_V=0$ and measure its overlap with later record energy
configurations in valley $n_V=k$. This allows us to define a
correlation between $c(k)=c(n_V=0;n_V=k)$ via the Hamming distance
$H(k)$ between them:
\begin{eqnarray}
c(k)=1-\frac{H(k)}{N/2}.
\label{correq}
\end{eqnarray}
Fig.~\ref{tau_corr} shows that the extremal search decorrelates
uniformly faster for decreasing $\tau$, similar to a thermal search
for increasing $T$. But for $\tau>1$ the correlations decay about
exponentially with the valley index and appear to converge to a
plateau value $c_{\infty}$ at large $k$, similar to a low-temperature
search. Such a plateau indicates a certain number of ``frozen''
variables which provide a ``backbone'' for any near-optimal
configuration. In contrast, for $\tau<1$ the extremal search
decorrelates almost instantly, and there is only a plateau at
$H=N/2$, the most likely distance for any two random
configurations. 

\begin{figure}
\vskip 2.3in \includegraphics{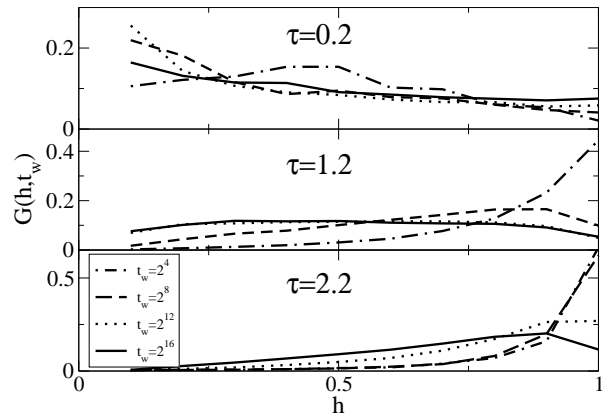}
\caption{Plot of the hitting probability $G(h,t_w)$ as a function of
  the hitting time $h$ defined in Eq.~(\protect\ref{hittingeq}) for
  valleys with increasing entry times $t_w$. Shown are the
  probabilities for $\tau$-EO for $\tau=0.2$, 1.2, and 2.2, all at
  $L=16$. Note that for $\tau=1.2$ a near-uniform distribution is
  quickly reached while for small and large $\tau$ small and large
  hitting times $h$ predominate, resp.}
\label{hittingplot}
\end{figure}

Finally, we also look at the hitting time for energy records with a
valley, which illuminates their internal structure. If we call $t_w$
(``waiting time'') the time when a valley was entered, $t_m$ the time
when the energy minimum was hit, and $t_x$ the time when the valley
was exited (and the next one entered), then Ref.~\cite{Dall03} defined
a relative hitting time as
\begin{eqnarray}
h=\frac{t_m-t_w}{t_x-t_w}, \quad 0 \leq h \leq 1,
\label{hittingeq}
\end{eqnarray}
(called ``$\tau$'' in Ref.~\cite{Dall03}). For valleys without any
structure, the average $h$ would be close to zero, but close to one
for highly structured valleys. In Ref.~\cite{Dall03} for the thermal
search at low temperatures it was found that the distribution for $h$
is peaked at large $h$ values, indicative of the high internal
structure of the spin glass valleys.  In Fig.~\ref{hittingplot}, we
plot the probability density function $G(h,t_w)$ of observing hitting
time $h$ in a valley entered at $t_w$.  We find a strong variation
with the parameter $\tau$ in the way $\tau$-EO explores that internal
structure. For large $\tau$, the extremal search behaves similar to
the lowest-temperature thermal search, i.e.\ $G$ is right-skewed,
while for very small $\tau$, the extremal search seems to ignore the
internal structure and discovers what it considers the minimum quite
quickly. For the search at $\tau=1.2$, which typically provides the
best energy overall, $h$ is almost uniformly distributed over the time
spent in the valley, at least for large $t_w$.

\begin{figure}
\vskip 2.3in \includegraphics{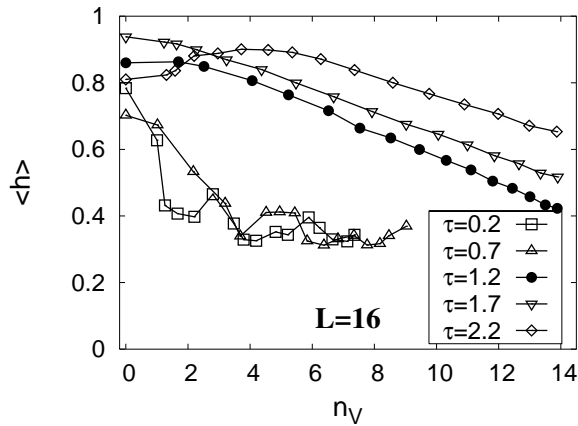}
\caption{Plot of the average hitting time $\langle h\rangle$ defined
in Eq.~(\protect\ref{hittingeq}) as a function of the valley index
$n_V$. For too small $\tau$, extremal search quickly saturates at
short hitting times, unable to leave the current valley. There is a
noticeable transition in behavior for $\tau>1$ with a linear decay in
$\langle h\rangle$ with a slope that slowly decreases with increasing
$\tau$.  }
\label{tau_hitting}
\end{figure}

In Fig.~\ref{tau_hitting}, we find that the average hitting time
$\langle h\rangle$ decreases for increasing valley index, ever more
rapidly for smaller $\tau$, until it saturates. It may appear that
hitting onto the record minimum soon after entering a valley may be a
good thing for an optimization with a local search. Yet, since $h$ is
measured relative to the length of the residence within a valley,
small $\langle h\rangle$ here means only that it takes a long time to
exit a valley, without ever taking full account of its internal
structure. The best compromise in terms of finding energy records (and
leaving valleys) quickly seems to be provided by $\tau=1.2$.

\subsection{Varying System Size}
\label{sizevary}
Based on the findings in Sec.~\ref{tauvary}, it appears that the most
favorable behavior of $\tau$-EO from the standpoint of optimization is
obtained near $\tau=1.2$. To study the behavior of the extremal search
for varying system size in comparison with the thermal search in
Ref.~\cite{Dall03}, we will therefore fix $\tau=1.2$ in this section.

\begin{figure}
\vskip 2.3in \includegraphics{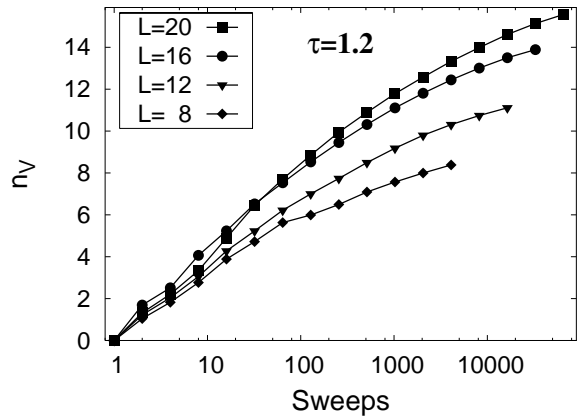}
\caption{Plot of the valleys found by EO on a logarithmic timescale at
fixed $\tau=1.2$ but for system sizes $L=8$, 12, 16, and 20. Progress
toward new valleys occurs linearly on a logarithmic time scale,
altered at later times by finite size effects, which become less
pronounced for larger $L$. }
\label{nV}
\end{figure}

First, we look again at the number of valleys found for increasing
runtime. As Fig.~\ref{nV} shows, the gain in the number of valleys
entered increases roughly on a logarithmic timescale. The growth slows
at later times, apparently due to system size effects, since the
increase becomes more linear, and differences between data less
pronounced, for increasing system size. These findings are very
similar to fixed-temperature data (see inset of Fig.~1) in
Ref.~\cite{Dall03}.

\begin{figure}
\vskip 2.3in \includegraphics{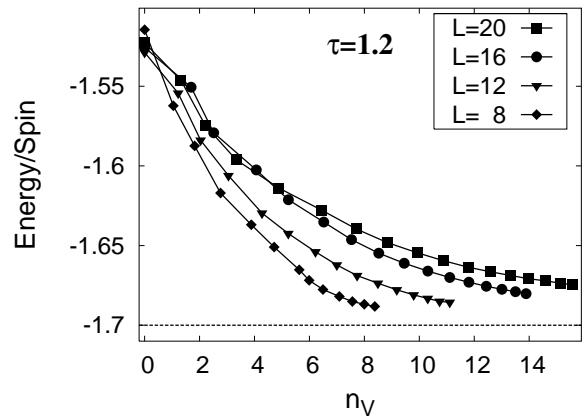}
\caption{Plot of the energy records found by EO within a given valley
as a function of the valley index $n_V$. Here $\tau=1.2$ is fixed and
system size $L$ is varied. The dashed line at $\approx-1.70$ marks the
average ground state energy density for
$L\to\infty$~\protect\cite{Pal96}. Convergence toward this limit at
our fixed runtime becomes ever more difficult for increasing $L$.
Yet, the saturation at larger $n_V$ clearly indicates the proximity of
the ground state effecting the search.}
\label{energy}
\end{figure}

After one system sweep from a random initial state, the energy per
spin reaches a typical level that strongly depends on $\tau$, as
Fig.~\ref{tau_energy} shows, albeit very little on $L$, according to
Fig.~\ref{energy}. But progressing further toward lower energy records
through subsequent valleys soon yields diminishing returns. In
contrast, for the thermal search in Ref.~\cite{Dall03} (see Fig.~4
there) new energy records provide a constant gain $\Delta$ in energy,
which in itself depends on $L$ and $T$. It should be noted, though,
that the extremal search reaches extremely low energies quickly and
becomes sensitive to the presence of the ground state. The behavior of
a thermal algorithm is more closely resembled at smaller $\tau$, such
as $\tau=0.7$ in Fig.~\ref{tau_energy}, for which the decrease in
energy scales linearly with $n_V$.

\begin{figure}
\vskip 2.3in \includegraphics{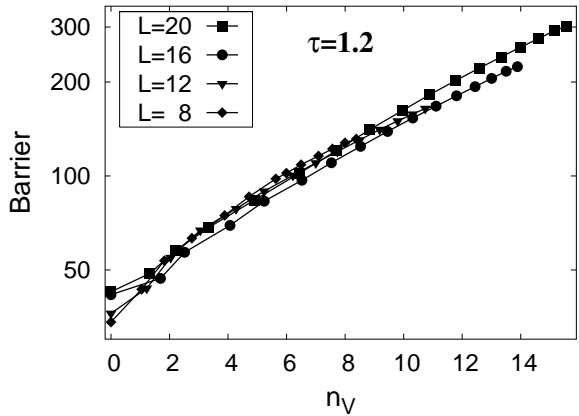}
\caption{Plot of the barrier height as a function of valley index
$n_V$ as in Fig.~\protect\ref{tau_barrier} but for fixed $\tau=1.2$
and different system sizes $L$. Since $\tau>1$, barrier heights rise
exponentially. Barriers are measured relative to the most recent
low-energy record and appear to be independent of $L$. }
\label{barrier}
\end{figure}

As in Fig.~\ref{tau_barrier} for $\tau>1$, Fig.~\ref{barrier} shows
that barrier heights scale exponentially for $\tau=1.2$, independent
of system size. In fact, that data for different $L$ appears to
collapse automatically, without any rescaling. This effect can be
explained by the definition of barrier heights as measured {\it
relative} to the lowest preceding energy record. This would imply that
to leave the $i$-th valley an extremal search needs to scale a certain
barrier height that is largely insensitive to the system size. (Any
deviation from collapse could well be due to a certain arbitrariness
in gauging $n_V=0$ after a single sweep.) Although on a different
scale, the barrier heights for a thermal search were also found to be
only weakly dependent on system size (see Fig.~2 in
Ref.~\cite{Dall03}).

\begin{figure}
\vskip 2.3in \includegraphics{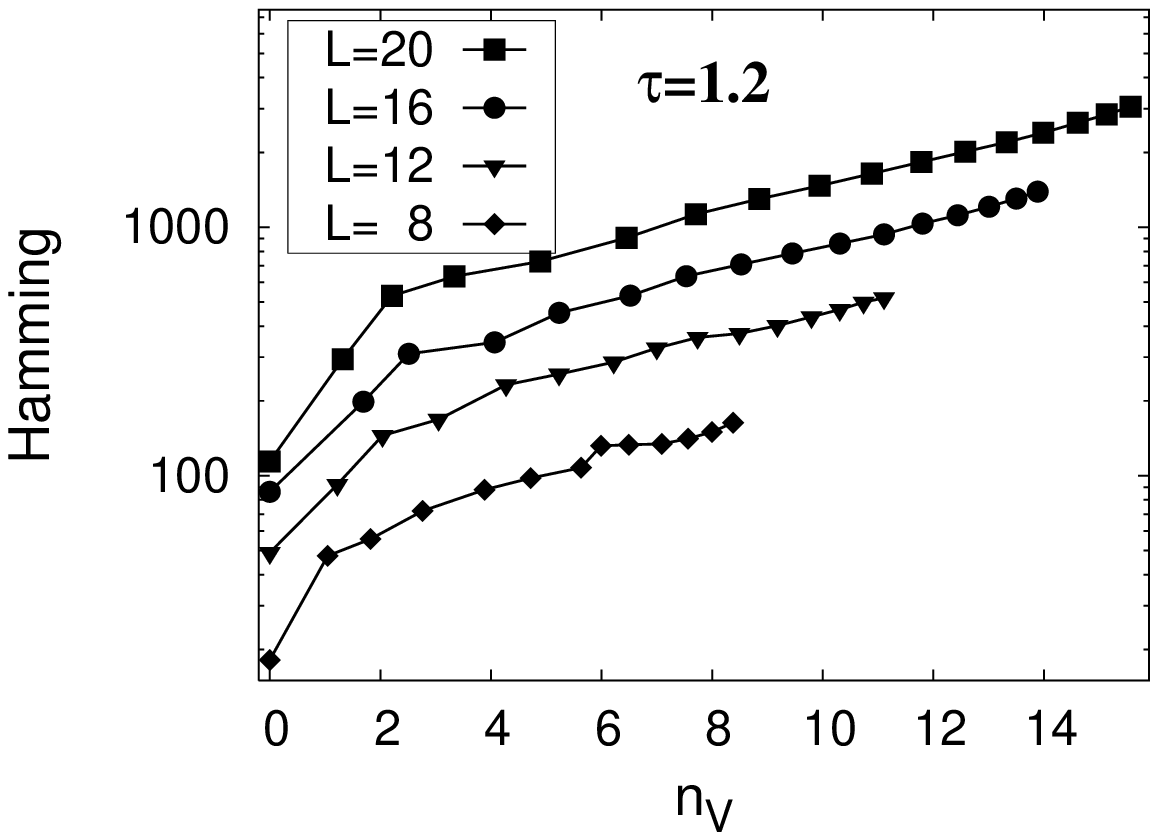} \includegraphics{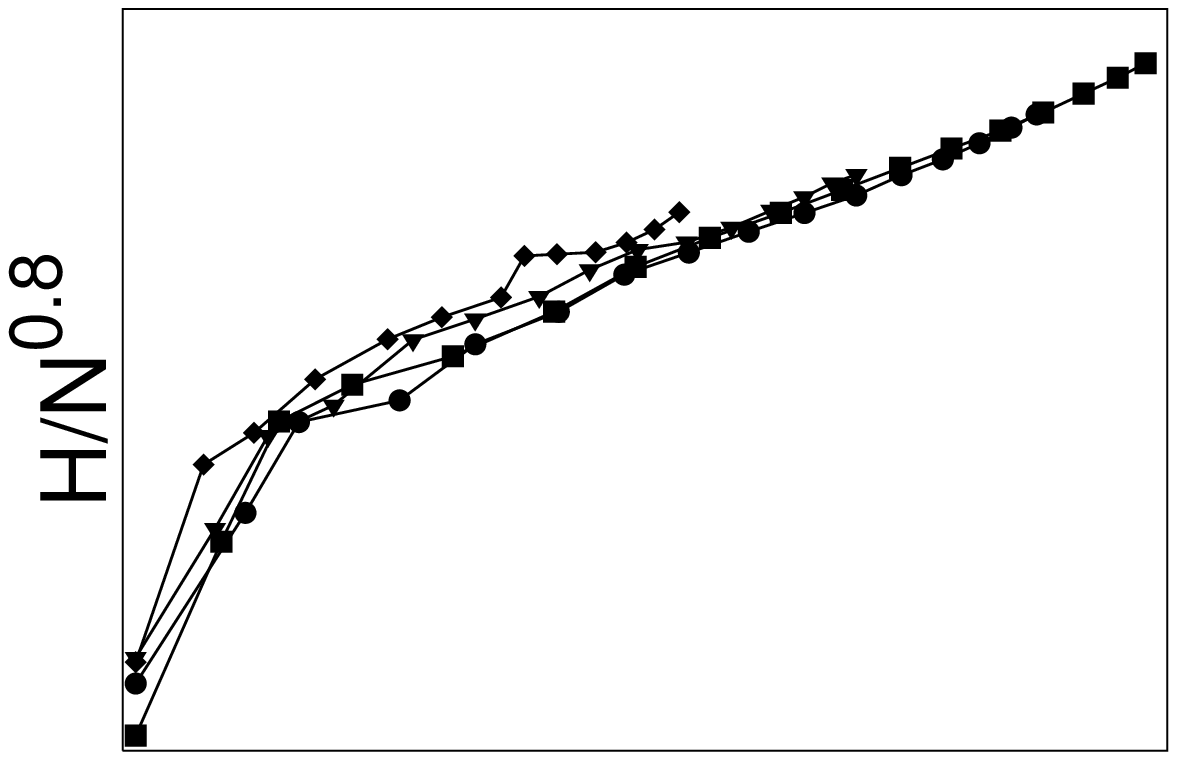}
\caption{Plot of the Hamming distance between successive low-energy
records as a function of valley index $n_V$ at fixed $\tau=1.2$ and
variable $L$. The inset shows the same data rescaled by $N^{0.8}$
where $N=L^3$. }
\label{hamming}
\end{figure}

Hamming distances between consecutive energy records are shown in
Fig.~\ref{hamming}. Similar to the barrier height, the Hamming
distance also grows exponentially with the valley index $n_V$, showing
a significant, but non-extensive, $L$-dependence.  The Hamming
distances found for $\tau=1.2$ here correspond about to the largest
found in the thermal search in Ref.~\cite{Dall03} at the highest
temperatures, {\it i. e.} just below $T_g$. This demonstrates the
breadth of the extremal search for an optimal choice of $\tau$.

Eliminating the valley-index dependence between barrier heights in
Fig.~\ref{barrier} and Hamming distances in Fig.~\ref{hamming}, as in
Fig.~\ref{tau_HamBar}, we again find a nearly linear relation. This
fact re-affirms the purely geometrical origin of this relation,
independent of system size.

\begin{figure}
\vskip 2.3in \includegraphics{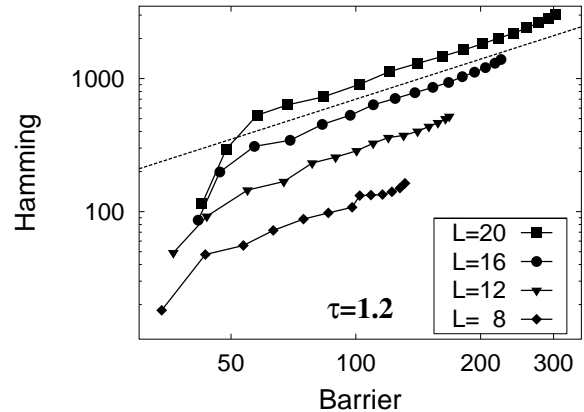}
\caption{Plot of the Hamming distance as a function of the barrier
height at fixed $\tau=1.2$ and variable $L$. This data was obtained
from Figs.~\protect\ref{barrier} and~\protect\ref{hamming} by
eliminating the valley-index dependence between them. Linearity is
exemplified by the dashed line of slope 1. }
\label{HamBar}
\end{figure}

\begin{figure}
\vskip 2.3in \includegraphics{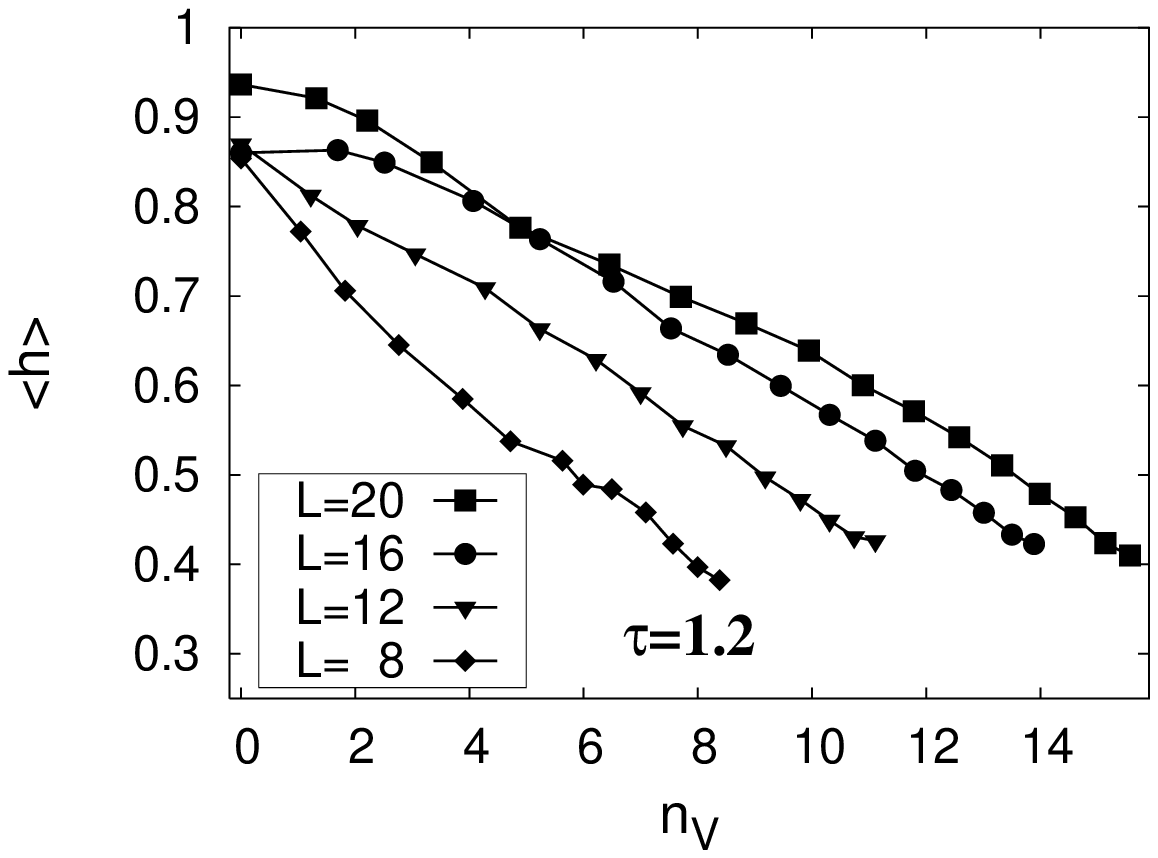} \includegraphics{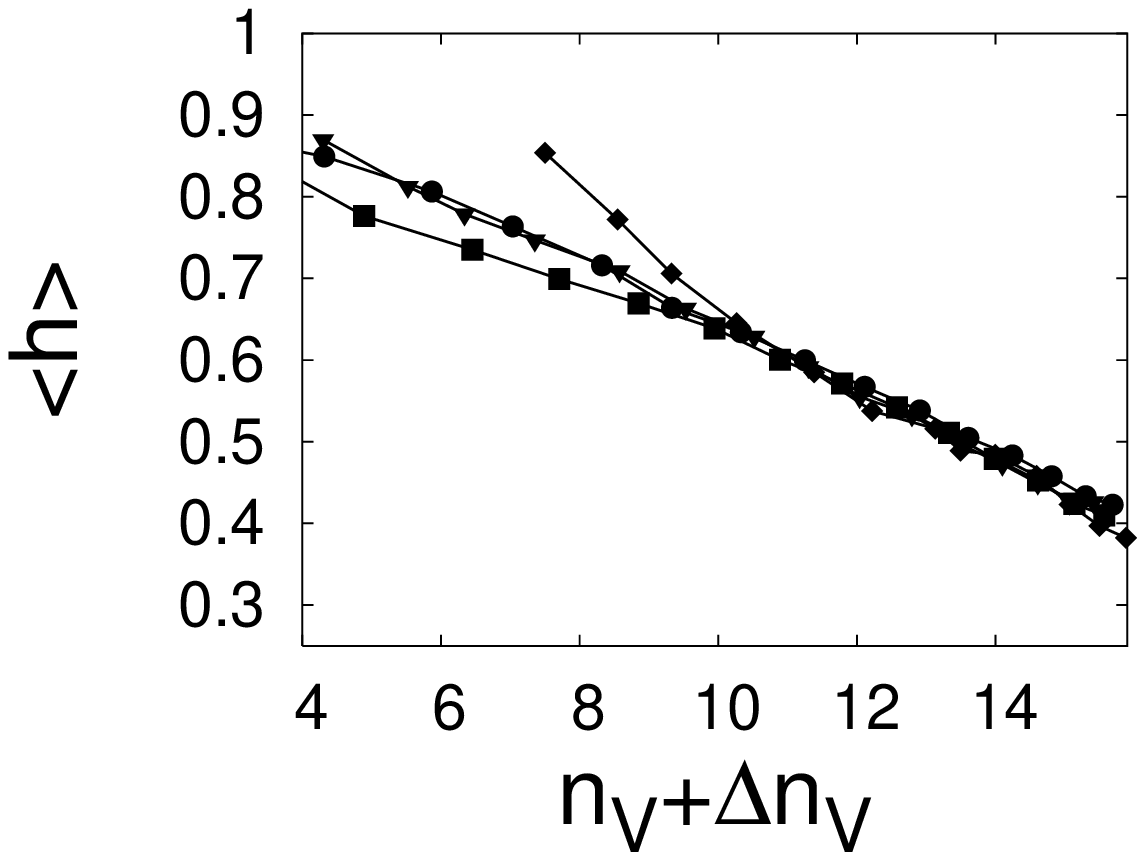}
\caption{Plot of the average hitting time $\langle h\rangle$ defined
in Eq.~(\protect\ref{hittingeq}) as a function of the valley index
$n_V$ for fixed $\tau=1.2$ and various $L$. Like in
Fig.~\protect\ref{tau_hitting}, $\langle h\rangle$ falls linearly with
$n_V$, but with a $L$-independent slope. The inset shows that a simple
shift in $n_V$ collapses the data for large $n_V$ quite well.  }
\label{hitting}
\end{figure}

{}Finally, in Fig.~\ref{hitting} we show the behavior of the average
hitting time $\langle h\rangle$ for different $L$. In each case,
$\langle h\rangle$ falls linearly with valley index $n_V$ with a
size-independent slope. Hence, the data can be collapsed with a simple
shift in $n_V$. That shift is not uniform in $L$, which could well be
due to the ambiguity in gauging $n_V$.

\section{Conclusions}
\label{conclusions}
 The energy landscape of a spin-glass is probed in this work with a
non-thermal optimization algorithm, Extremal Optimization (EO).
Energy values of record magnitude are used to partition the states
visited into a sequence of valleys, indexed by $n_V$.  Within each
valley the state of lowest energy, or minimum, is the best result
obtained with the algorithm on a given time scale.  The energy barrier
separating two consecutive minima is, by construction, the largest
ever barrier scaled on the same time scale.  The sole adjustable
parameter of EO, $\tau$ and the system size $N=L^3$ are both varied
systematically in the investigations.
  
Comparing our present results to those of Ref.~\cite{Dall03}, which
uses the same classification scheme in connection with a thermal
algorithm, the Waiting Time Algorithm (WTM), we are able to
disentangle the intrinsic, or geometrical, properties of the landscape
from those specifically linked to the two algorithms.  These aspects
are separately discussed in this sequel.
 
Both EO and WTM uncover a non-trivial valley structure whose overall
features are broadly similar.  The first observation is that, in order
to find states of lower energy, progressively higher barriers must be
surmounted.  The extremal property of these intervening barriers
implies that the sequence of lowest minima encountered can
(approximately) be treated as a Poisson process with logarithmic time
arguments~\cite{Dall03}. Such description can only work as long as the
system remains far away from the global minimum. Indeed, the number
$n_V$ of valleys explored grows on average in near logarithmic fashion
and the decreasing logarithmic slope apparent for the 'best' value of
$\tau$ likely reflects the fact that once the system is near to the
global optimum, further improvements become harder to achieve.

The magnitude of the highest barrier scaled grows linearly with $n_V$
in the WTM analysis, but exponentially with EO. The same is true for
the Hamming distance between consecutive minima. By eliminating $n_V$,
from these exponential relationships a linear dependence emerges,
which links the barrier and the Hamming distance between consecutive
minima. This concurs with the WTM result, and the linear dependence
thus stands out as a geometric property of the energy landscape.  For
growing system size, and $\tau = 1.2$, the Hamming distance between
consecutive minima grows as $N^{0.8}$, which is qualitatively similar
to the $N^{0.95}$ scaling found with the WTM. Also similar is the
decay of the configuration overlap between minima $k$ valleys apart,
in both cases nearly exponential in $k$.

The aspect where WTM and EO mostly differ is in the distribution of
the time spent searching for the lowest energy state within a fixed
valley.  Unlike the WTM, EO locates this minimum fairly quickly, at
least for values of $\tau$ close to $1.2$, which is the best value in
terms of optimization performance. In other words, EO seems to be able
to roam more easily through configuration space, which is in accord
with the original intention behind its design.  Short-ranged spin
glasses are probably the test problem where EO is at its worst in
terms of optimization performance, yet its performance is slightly
better than what WTM can achieve in terms of, say, the lowest energy
achieved in a fixed number of updates. However, from a broader
optimization perspective the performance of the two algorithms is
similar, since lower energy value are logarithmically spaced in
time. This, we suggest, might be a general feature of local searches
in complex landscapes with a large number of near-equivalent minima.

{}From a physics point of view, applying a non-thermal algorithm to
landscape exploration removes the appeal of using concepts adapted
from thermal-equilibrium, such as the free energy, to describe the
dynamics.  Since, as we have shown, the process of jumping from one
valley to a lower-lying valley, is only weakly dependent on the
algorithm chosen, thermal concepts are likely to be generally
irrelevant for the \emph{drift} part of the dynamics even though the
dynamical update rules obey detailed balance. As we have argued
elsewhere~\cite{Sibani03,Sibani04} the reason for this is that these
jumps are effectively irreversible on the time scale at which they
occur.

\section{Acknowledgments}
We would like to thank J. Dall for helpful discussions. This work was
partially funded by NSF grant DMR-0312510.



\begin{thebibliography}{999}

\bibitem{Hartmann04}
{\em New Optimization Algorithms in Physics,} 
eds. A. K. Hartmann and H. Rieger, (Wiley-VCH, Weinheim, 2004).

\bibitem{Frauenfelder96} 
{\em Landscape Paradigms in Physics and Biology,}
eds. H. Frauenfelder et. al. (Elsevier, Amsterdam, 1997).

\bibitem{Boettcher00}  S. Boettcher and A. G. Percus,
Artificial Intelligence {\bf 119}, 275 (2000).

\bibitem{Boettcher01a}  S. Boettcher and A. G. Percus,
Phys. Rev. Lett.  {\bf 86}, 5211 (2001).

\bibitem{Dall03}
J. Dall and P. Sibani,
Eur.~Phys.~J.~B {\bf 36}, 233-243 (2003).

\bibitem{Nemoto88} 
 K. Nemoto,
J. Phys. A: Math. Gen. {\bf 21}, L287 (1988).
 
 D. Vertechi and M.A. Virasoro, 
J. Phys. France {\bf 50}, 2325-2332 (1989). 

\bibitem{Vertechi89} 
 D. Vertechi and M.A. Virasoro, 
J. Phys. France {\bf 50}, 2325-2332 (1989).  
  
\bibitem{Billoire01}
A. Billoire and E. Marinari,
J. Phys. A: Math. Gen. {\bf 34}, L727-L734 (2001).

\bibitem{Sibani89}
P. Sibani and K.~H. Hoffmann,
Phys. Rev. Lett. {\bf 63}, 2853 (1989).

\bibitem{Lederman91}
M. Lederman, R. Orbach, J.M. Hammann,  M. Ocio and E. Vincent,
Phys. Rev. B, {\bf 44}, 7403 (1991)
 
\bibitem{Sibani91}
P. Sibani and K. H. Hoffmann,
Europhys. Lett. {\bf 16}, 423 (1991).

 \bibitem{Bouchaud95}
 J. P. Bouchaud and D.S. Dean, 
 J. Phys. I France. {\bf 5}, 265 (1995).                                                             
\bibitem{Joh96}
 Y.~G. Joh, R. Orbach, and J. Hamman,
Phys. Rev. Lett. {\bf 77}, 4648 (1995).

\bibitem{Hoffmann97}
K. H. Hoffmann, S. Schubert, and P. Sibani,
Europhys. Lett. {\bf 38}, 613 (1997). 

\bibitem{Sibani97a} 
 P. Sibani and K. H.   Hoffmann,                                 
Physica A {\bf 234}, 751 (1997).                                                          
 
\bibitem{Joh99}
Y. G. Joh,  R. Orbach, G. G. Wood,  J. Hammann and E. Vincent,
Phys. Rev. Lett. {\bf 82}, 438 (1999).

\bibitem{Crisanti00} 
A.~Crisanti and F.~Ritort,
Europhys. Lett. {\bf 52}, 640 (2000).  

\bibitem{Buisson03}
L.~Buisson, L.~Bellon, and S.~Ciliberto,
J. Phys. Cond. Mat. {\bf 15}, S1163 (2003).

\bibitem{Kirkpatrick83} 
S. Kirkpatrick, C. D. Gelatt, and M. P. Vecchi,
Science {\bf 220}, 671 (1983).
 
\bibitem{Salamon02}
P. Salamon, P. Sibani, and R. Frost,
{\it Facts, Conjectures, and Improvements for Simulated Annealing}
(Society for Industrial \& Applied Mathematics, 2002).

\bibitem{Boettcher99}
S. Boettcher,
J. Math. Phys. A: Math. Gen. {\bf 32}, 5201-5211 (1999).

\bibitem{Boettcher04} S. Boettcher and A. G. Percus,
Physical Review E {\bf 69}, 066703 (2004).

\bibitem{Boettcher03}
S. Boettcher,
Eur.~Phys.~J.~B {\bf 31}, 29-39 (2003).

\bibitem{Boettcher02}
S. Boettcher and M. Grigni,
J. Phys. A. {\bf 35}, 1109 (2002).

\bibitem{Boettcher01b} S. Boettcher and A. G. Percus,
Phys. Rev. E {\bf 64}, 026114 (2001).


\bibitem{Middleton04}
A. A. Middleton,
Phys. Rev. E {\bf 69}, 055701 (R) (2004). 

\bibitem{Dall01}
J. Dall and P. Sibani,
Computer Physics Communication {\bf 141}, 260 (2001).


\bibitem{Pal96}
K. F. Pal,
Physica A {\bf 233}, 60-66 (1996).


\bibitem{Sibani98}
P. Sibani,
Physica A {\bf 258}, 249 (1998).

\bibitem{Sibani03} 
 P. Sibani  and J. Dall,
 Europhys. Lett.  {\bf 64}, 8-14 (2003). 

\bibitem{Sibani04} 
P. Sibani and H. J. Jensen,
Europhys. Lett. {\bf 69}, 563-569 (2005).
\end{thebibliography}
\end{document}